\documentclass[aps,prb,twocolumn,floatfix,superscriptaddress]{revtex4-1}
\usepackage{graphicx,epsfig,array}
\usepackage{color,graphicx,amssymb,amsmath}
\begin{document}

\title{ A variational approach for calculating Auger electron spectra:
going beyond the impurity approximation}

\author{Anamitra Mukherjee}
\affiliation{Department of Physics and Astronomy, University of British
Columbia, Vancouver, BC, Canada, V6T 1Z1}

\author{George A. Sawatzky} 
\affiliation{Department of Physics and Astronomy, University of British
Columbia, Vancouver, BC, Canada, V6T 1Z1}
\affiliation{Quantum Matter Institute, University of British Columbia,
Vancouver, BC, Canada, V6T 1Z4}
\affiliation{Department of Chemistry, University of British
Columbia, Vancouver, BC, Canada, V6T 1Z1}           

\author{Mona Berciu}

\affiliation{Department of Physics and Astronomy, University of British
Columbia, Vancouver, BC, Canada, V6T 1Z1}
\affiliation{Quantum Matter Institute, University of British Columbia,
Vancouver, BC, Canada, V6T 1Z4}

\date{\today}

\begin{abstract}
We propose a novel variational method to calculate the two-hole
propagators relevant for Auger spectroscopy  in transition metal
  oxides. This method can be thought of as an intermediary step
between the full solution (which is difficult to generalize to systems
with partially filled bands) and the impurity approximation. Like the
former, our solution has full translational invariance, and like the
latter, it can be generalized to  certain types of systems with partially filled
bands. Here we compare both our variational approximation and the
impurity approximation against the exact solution for a simple
one-dimensional model with filled bands.  We show that when the
  energies of the eigenstates residing primarily on the transition
  metal ions do not overlap with those of the eigenstates residing
  primarily on Oxygen ions, both approximations are valid but the
  variational approach is superior.
\end{abstract}
\maketitle

\section{Introduction}

Spectroscopic measurements are a powerful set of tools for probing
various aspects of many-body physics.\cite{ pes-gen,
chattarji-book} Among these, Auger electron spectroscopy (AES)
provides information about the local atomic multiplet structure,
on-site interaction strengths and the crystal fields.\cite{auger-exp-1,
auger-rev-0, auger-rev} In transition metal oxides, the AES of
  the transition element can be supplemented by  the O 
  \textit{ KLL} Auger spectra, resulting in additional information
  about the O on-site repulsion energy as well as  interactions
  between holes located on nearest-neighbor transition metal and O
  ions.\cite{oxy-corr} Such information is vital for understanding
correlated materials, which is why AES has been the subject of sustained research for a long time.

The Auger process consists of the decay of a core hole into two
  final state holes (initially located at the atomic site where the
  original core hole was created by the high energy X-ray) plus an
  Auger electron, and is mediated by onsite Coulomb interactions.
One of the most studied cases has these two final state holes residing
in the valence band, and goes by the name of Core-Valence-Valence
(CVV) Auger spectroscopy. From a theoretical point of view, the
easiest case to handle has a full valence band except for the two
Auger holes.  The central quantity of interest is the two hole Green's
function, which for an otherwise full band can be calculated within
the two-step approximation using the Cini-Sawatzky theory.\cite{cs-1,
  cs-2, cs-3} The resulting two hole spectral function, multiplied by
momentum dependent matrix element factors, provides the theoretical
predictions for AES.\cite{pothoff-sprectro-general} Many extensions
have been proposed to incorporate various aspects 
such as dynamical screening,\cite{dyn-scrn-1, dyn-scrn-2, dyn-scrn-3}
off-site interactions,\cite{off-site} overlap effects,\cite{overlap-1,
  overlap-2, overlap-3} and one step formulation.\cite{one-step} We
refer the reader to a recent review \cite{auger-prog} for more
details. These efforts have led to spectacular success in explaining
AES for materials such as Cu and Cu$_2$O.

\begin{figure}[b]
  \centering{
  \includegraphics[width=0.7\columnwidth]{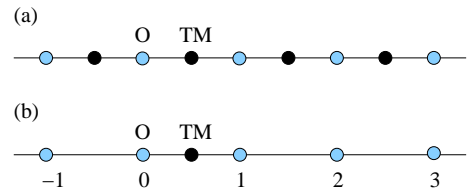}}
\vspace{-.0cm}
  \caption{
  (a) Sketch of the 1D periodic Anderson model. 
  'TM' and  'O' represent transition metal and O ions,
  respectively. (b) In the impurity approximation,  an impurity 'TM'
  ion is coupled to the  1D 'bath' of O.}
\vspace{-.0cm}
\label{f-1}
\end{figure}

However, the problem of understanding AES for systems with a partially
filled valence band, like the oxides of transition elements Ni, Co,
Fe, Mn, etc., remains open, because the two hole spectral function is
very challenging to compute in this case. This is because in the
  presence of other holes, the dynamics of the two additional holes is
  a complicated many-body problem, whereas in an otherwise full band
  the two holes interact only with one another (if we ignore
  electron-hole excitations between the valence and the conduction bands), i.e. this is a
  two-body problem.  Limited success has been achieved employing
variants of the bare ladder approximations\cite{ladder} and assuming
low hole density in the bands.  A completely different approach is to
use the Anderson impurity approximation, whose underlying idea is as
follows: in transition metal oxides, the transition metal atoms are
typically connected to each other through oxygen ligands. The simplest
example in one dimension is sketched in Fig.~\ref{f-1}(a), and has
transition metal atoms intercalated with O atoms. In the impurity
approximation,\cite{impurity-apr,ia2,ia3,ia4} the full problem is
simplified to that of a single transition metal atom coupled to the
bath of O, as sketched in Fig.~\ref{f-1}(b). This greatly simplifies
the calculation and is a reasonable step towards understanding local
multiplet structures. However, because the symmetry of the problem is
lowered, momentum-resolved spectral weights cannot be calculated.

Here, we propose a variational approach for finding the two-hole
Green's functions needed for AES, which can be thought of as an
intermediary step between the impurity approximation and the full
lattice calculation. Like the former, it has a reduced Hilbert space
and can therefore be generalized  to (some) systems with
partially filled bands. Like the latter,  it has  full
translational invariance so that momentum-resolved spectral weights can be
calculated. We argue that the location of the spectral weight
component ignored within our approximation can be inferred apriorily
but should be small for materials where AES is a
useful probe. Moreover, the variational space can be systematically
enlarged  to check the relevance of some of the neglected terms.

In this paper we present the underlying idea and the general
formalism of this variational 
approach, which is based on a recently developed method to calculate
many-particle Green's functions.\cite{cf-prl, cf-ext, Ashley} Here, we apply
it to the  simple model sketched in Fig. \ref{f-1} and assume otherwise
full bands, so that we can benchmark it against the 
exact solution available in this case. For completeness, we also show
impurity approximation results for this model. This allows us to gauge
the advantages and disadvantages of both these approximations and to
understand for what regions in the parameter space they are
valid. This information will guide us, in future 
work, to using this method for appropriate systems where the transition elements have
partially filled $d$ orbitals.

 The paper is organized as follows. In Sections II we define the
 Hamiltonians  and in Section III we discuss the
methods to calculate two-hole propagators exactly and with the two
approximations.  In Section IV we 
present the exact results and compare them with those predicted by our
variational approximation as well as the impurity approximation
for two different topologies of coupling between
the impurity and the bath of O. Section V contains the summary and
conclusions. Some of the details are 
presented in the two Appendixes.

\section{Hamiltonian}

\subsection{The periodic Anderson model} 

The periodic 1D Anderson model shown in
Fig.~\ref{f-1}(a), is defined by the Hamiltonian:
$$ {\cal H_{P}} = {\cal H}_{\rm TM} + {\cal H}_{\rm O} + {\cal H}_{\rm
  hyb}
$$ where ${\cal H}_{\rm TM}$ describes the  TM atoms, which for
simplicity are assumed to have only two (spin-degenerate) states each,
and is given by:
$$ {\cal H}_{TM} = U \sum_{i}^{} n_{d,i,\uparrow} n_{d,i,\downarrow} +
\Delta \sum_{i,\sigma}^{} n_{d,i,\sigma}.
$$  The O
``bath'' is described by a 1D Hubbard model: 
$${\cal H}_{\rm
  O}=-t\sum\limits_{i,\sigma}(c^{\dagger}_{i,\sigma}c_{i+1,\sigma} +
h.c.)  + U_0\sum\limits_{i} n_{i,\uparrow}n_{i,\downarrow}
$$ while the TM-O hybridization is described by
$$ {\cal H}_{hyb} = - V \sum_{i,\sigma}\left[d^\dagger_{i,\sigma}
  \left( c_{i,\sigma} + c_{i-1,\sigma}\right) + h.c.\right].
$$   Here, all creation
operators are electron creation operators, with $c_{i\sigma}$ and
$d_{i,\sigma}$ the operators for  O
and TM orbitals, respectively, with the convention that the i$^{th}$ TM atom
is placed to the left of the i$^{th}$ O atom. As usual,
$n_{d,i,\sigma} = d^\dagger_{i\sigma} d_{i\sigma},
n_{i\sigma}=c^\dagger_{i\sigma} c_{i\sigma}$. For book-keeping purposes
we assume that there are $N$ TM and $N$ O sites, but in the final
calculation we let $N\rightarrow \infty$. Because the method we
  introduce below is based on a 
  real-space representation, addition of longer-range interactions,
  such as repulsion between holes residing on
  neighboring TM and O sites, is trivial to implement. This model
assumes that only the sigma-bonding
O $2p$ orbitals pointing towards the TM neighbors are relevant. 
 Generalization to models that include more O and/or  TM
orbitals, as well as lattices in higher
dimensions, is discussed below.

For AES in a system with full bands, we start from  the completely
full ground state
$|\Omega\rangle_P=\prod_{i,\sigma}^{}d^{\dagger}_{i, \sigma}
c^\dagger_{i,\sigma}|0\rangle $ of energy $E^P_\Omega=N(2\Delta + U
+U_0)$,  and create two holes. Because the periodic Hamiltonian is invariant to
translations and because we want to describe states where the two
holes can be at the same site,  we choose 
two-hole basis states with a total momentum $k$ and zero spin
projection,\cite{cf-prl} namely 
$|k,n,dd\rangle=\frac{1}{\sqrt{N}}\sum_ie^{ik(R_i+na/2)}d_{
  i, \uparrow}d_{ i+n, \downarrow}|\Omega\rangle_P$ if both holes
are on TM sites,
$|k,n,dc\rangle=\frac{1}{\sqrt{N}}\sum_ie^{ik(R_i+na/2)}d_{i,
  \uparrow}c_{ i+n, \downarrow}|\Omega\rangle_P$ 
and 
$|k,n,cd\rangle=\frac{1}{\sqrt{N}}\sum_ie^{ik(R_i+na/2)}c_{i, 
   \uparrow}d_{ i+n, \downarrow}|\Omega\rangle_P$ 
if one hole
is on a TM site and the other is at an O site, and finally
$|k,n,cc\rangle=\frac{1}{\sqrt{N}}\sum_ie^{ik(R_i+na/2)}c_{i,
  \uparrow}c_{ i+n,\downarrow}|\Omega\rangle_P$ if both holes are in
the O bath. Here $a$ is the lattice constant and $n = -{N\over2}
+1,\dots, {N\over 2}$ takes all possible values consistent with the
periodic boundary conditions. Taken together, these
states constitute a full basis for 
the Hilbert subspace containing states with total momentum $k$ and
zero spin projection. 

The
aim is to find the propagator
$$G_{dd}(k,0,\omega)=\langle
k,0,dd|\hat{G}_P(\omega)|k,0,dd\rangle$$ where
$\hat{G}_P(\omega)=[\omega+i\eta-({\cal 
    H_P}-E^P_\Omega)]^{-1}$ is the resolvent for
${\cal H}_P$, because its associated spectral weight
$A_{{dd}}(k,0,\omega)=-\frac{1}{\pi}\mbox{Im}[G_{{dd}}(k,0,\omega)]$,
which has poles at energies $\omega = 
E_{2h}(k)-E^P_\Omega$ for any two-hole eigenstate with
total momentum $k$ and energy $E_{2h}(k)$, is proportional to the
momentum-resolved  spectral intensity measured by AES. As discussed
below, our solution also provides the values of many other propagators 
beside $G_{dd}(k,0,\omega)$, from which other
useful information can be gleaned.

\subsection{Anderson impurity problem} 

The impurity approximation is a variational
  approximation where the holes are not allowed on any TM ions apart from
  the original one; this is equivalent with excluding all such
  orbitals from the variational space.
As a result, the full periodic problem is reduced to the Anderson impurity problem
sketched in Fig.~\ref{f-1}(b), and its Hamiltonian becomes:
$$ {\cal H}_I = {\cal H}_{\rm TM,I} + {\cal H}_{\rm O} + {\cal H}_{\rm
  hyb,I}
$$ where
$$ {\cal H}_{\rm TM,I}=U n_{d,\uparrow}n_{d,\downarrow}+\Delta
\sum\limits_{\sigma} n_{d,\sigma}
 $$ describes the impurity TM site, the O bath is described by ${\cal
  H}_{\rm O}$ as before, and
$$ H_{\rm hyb,I}=-V\sum\limits_{\sigma}
           [d^{\dagger}_{\sigma}c_{1,\sigma}+d^{\dagger}_{\sigma}
             c_{0,\sigma}+ h.c.]
$$ is their hybridization, with the impurity taken to be located
           between the bath O labeled '0' and '1'.  Here, again, all
           operators are electron operators and $d^{\dagger}_{\sigma}$ is
           the creation operator for the orbital of the TM impurity
           site.

The filled-band ground state is now $|\Omega\rangle_I=
\prod_{\sigma}^{}d^{\dagger}_{\sigma}\prod_{i}^{}
c^\dagger_{i,\sigma}|0\rangle$ and its corresponding energy is
$E^I_\Omega = 2\Delta + U + NU_0$.  To calculate AES-relevant spectra
we consider two-hole excitations in this ground state by removing two
electrons with opposite spins. Since invariance to translations is lost, the generic real
space states of interest are now $|{i\sigma}, {j\sigma'}\rangle \equiv
c^{}_{i\sigma}c^{}_{j\sigma'} |\Omega\rangle_I$, $|d\sigma, i
\sigma'\rangle = d_{\sigma} c^{}_{i\sigma'} |\Omega\rangle_I$ and
$|{dd}\rangle = d_{\uparrow} d_{\downarrow}|\Omega\rangle_I$.  Now we
are primarily interested in calculating the two-hole impurity Green's
function $G_{{dd}}(\omega)=\langle dd|\hat{G}_I(\omega)|dd\rangle$
where $\hat{G}_I(\omega)= [\omega+i\eta-({\cal
    H_I}-E^I_\Omega)]^{-1}$, and its corresponding two-hole spectral
function $A_{dd}(\omega)=-\frac{1}{\pi}\mbox{Im}[G_{dd}(\omega)]$.

\section{Method}

We begin with the exact solution for the periodic Anderson model.
To find the propagator of interest to us,
$G_{dd}(k,0,\omega)$, we generate its equation of
motion (EOM) from the identity: $\hat{G}_P(\omega)(\omega -{\cal
  H}_P+ E^P_\Omega+i\eta)=\hat{I}$. Calculating its
diagonal matrix element for $|k,0,dd\rangle$, we find:
\begin{widetext}
$$
(\omega  +2\Delta +U + i \eta)G_{dd}(k,0,\omega)
  = 1 + V\left[e^{ika\over 2}G_{cd}(k,1,\omega) +
    G_{cd}(k,0,\omega)+ e^{ika\over 2}G_{dc}(k,-1,\omega) +
    G_{dc}(k,0,\omega)\right] 
$$
\end{widetext}
where $G_{\alpha\beta} (k,n,\omega) =
\langle k,0, dd| \hat{G}_P(\omega) | k, n,
\alpha \beta \rangle$ for $\alpha,\beta = c, d$. In other words,
because 
the Hamiltonian links the state $|k,0,dd\rangle$ to states with one
hole on a TM site and one on a neighboring O, the EOM links
$G_{dd}(k,0,\omega)$ to propagators corresponding to such
states. Their EOM can be generated similarly, and we obtain an
infinite sequence of coupled linear equations. 

To solve it, we couple the propagators with holes at the same distance
$n$ in a vector:
$$
V_n=\left(\begin{array}{c}G_{dd}(k,n,\omega)\\G_{dc}(k,n,\omega)\\G_{cd}(k,n,\omega)
  \\ G_{cc}(k,n,\omega)\end{array}\right)
$$
and note that for any $n\ne 0$, the EOM can be grouped in the simple
recurrence relation:
$$ \gamma_n
V_n=\beta_nV_{n+1}+\alpha_nV_{n-1}
$$ for any given $k$ and $\omega$. Here $\gamma_n$, $\beta_n$ and
$\alpha_n$ are simple $4\times 4$  matrices that are read off
directly from the EOM. We
note that one can always  group the EOM in such simple recurrence
relations, even for models which allow longer-range
hopping\cite{cf-ext} and/or in higher dimensions.\cite{Ashley}
For $n=0$,  the recurrence relation also has an
inhomogeneous term:
$$ \gamma_{0}V_0=X+ \beta_{0} V_1 +
\alpha_{0} V_{-1}, $$ 
where $X^T=(1, 0, 0, 0)$ for this problem. The solution of such recurrence relations
has been discussed extensively elsewhere.\cite{cf-prl,cf-ext,Ashley} Briefly,
 we must have $V_n \rightarrow 0$ as $|n| \rightarrow
\infty$, because the Fourier transform of these propagators are the
amplitudes of probability to have the two holes  evolve from being on
the same TM site to being $n$ sites away from each other, in a given
time. As $|n|\rightarrow \infty$ this becomes very unlikely, and the
presence of the broadening $\eta$ which introduces an artificial
lifetime $1/\eta$ makes it even less so. As a result, for $n\ge 1$ we
have  $V_n=A_n(k,\omega)V_{n-1}$ where
$A_n=[\gamma_n-\beta_nA_{n+1}]^{-1}\alpha_n$ is calculated starting
with $A_{M}=0$ for a sufficiently large cutoff $M$. Similarly, for
$n\le -1$ we have $V_n=B_n(k,\omega)V_{n+1}$ where
$B_n=[\gamma_n-\alpha_nA_{n-1}]^{-1}\beta_n$ is calculated starting
with $B_{-|M|}=0$. Using $V_1=A_1
V_0$ and $V_{-1}=B_{-1}V_0$ in the $n=0$ equation gives:
$$ V_0 = \left[\gamma_0 - \beta_0 A_1(k,\omega) -
  \alpha_0B_{-1}(k,\omega)\right]^{-1} X.
$$ This gives us $G_{dd}(k,0,\omega)$ as the top 
entry in $V_0$. All other $n=0$ propagators, as well as those with $|n|
< M$, can also be then calculated efficiently. Projecting on a
different state than $\langle{k,0,dd}|$
simply requires using a different $X$, so other propagators can 
be found easily.

In principle, this method generalizes straightforwardly to lattices in
any dimension and with any topology, so no approximations should be
necessary. In practice, however, the computational cost quickly
becomes prohibitive. For instance, for models with nearest-neighbor
hopping in higher dimensions, one must group together in $V_n$ all
propagators where the holes are separated by $n_x a\hat{e}_x + n_y
a\hat{e}_y+\dots$ with $|n_x|+|n_y|+\dots = n$, i.e. with the same
Manhattan distance.\cite{Ashley} As a result, the dimension of $V_n$
increase roughly like $n^{d-1}$.  Adding more orbitals at the
TM/O sites will further amplify the problem, by increasing in a
combinatorial fashion the number of possible propagators with the same
$(n_x,n_y,\dots)$ separation.  As a result, while the solution can
still be cast in terms of continued fractions of matrices, their
dimensions increase fast with $n$ resulting in significant
computational costs. This is why efficient approximations
are needed.

As already mentioned, one much-employed option is the impurity
approximation. Its solution for $G_{dd}(\omega)$ can also be
formulated in terms of 
continued fractions of matrices (we present the details in the
Appendix, since we are not aware of a prior similar solution for this
problem). This approximation 
reduces the number of possible propagators by removing all but one TM
site from the problem. As a result, there is now only one state with
both holes at the impurity TM site as opposed to $N^2$ in the full
periodic problem (i.e., including all allowed total momenta), and only
$2N$ combinations with one hole at the TM and one at an O site, as
opposed to $2N^2$ options in the full problem. The number of states
associated with both holes in the O bath is not changed. 

The total number of distinct propagators is therefore reduced from
$4N^2$ in the full problem to $(N+1)^2$ in the impurity limit,
suggesting that the latter is considerably more efficient. However,
the full problem can be solved individually for each of the allowed
$N$ total momenta, since the translational invariance guarantees that
propagators with different momenta do not mix. Thus, one needs to
solve $N$ distinct problems with $4N$ propagators each, as discussed
above. In the impurity problem, loss of translational invariance means
that all propagators are coupled to one another through the EOM. From
this perspective, it is far less clear that using the impurity
approximation is computationally more efficient, although to fully
settle this one needs to take into account all other possible symmetries
and the effects of the truncation (for large system, the cutoff $M \ll
N/2$). In practice the O bath is often replaced by a
featureless density of states (we do not make this further
approximation in our calculations). This
certainly makes the impurity 
problem very efficient, but it also looses all information regarding the
role of the O bath topology, on top of the loss of
ability to momentum-resolve the AES spectral weights.

From this analysis, it is clear that a better strategy would be to
lower the total number of propagators while maintaining translational
invariance. This is the basis for our proposed variational method. Our
proposal is to remove the propagators which have the holes at different TM
sites. This is equivalent with excluding these states from the Hilbert
space, which is why this is a variational
approximation. Mathematically, this is easily achieved by setting
$G_{dd}(k,n,\omega)\equiv 0$ for all $n\ne 0$ in the EOM discussed
above, leading to vectors $V_n$, $n\ne0$ of dimension 3 instead of 4. Of
course, the saving is not big for this simple case, but it 
becomes considerable in higher dimensions and/or for more TM
orbitals. Because its variational space is significantly larger
  than that of the impurity approximation, this method is also
  guaranteed to be more accurate.

Physically, what this means is that both holes start at the same (any)
TM site. Eventually both hop into the O bath and move through the
system, but whenever they happen to both hop back to TM sites they
must go to the same TM ion. For systems with strong correlations,
i.e. where the states with the holes at the same TM site are at quite
different energies from states with the holes as different TM sites,
this should be a reasonable approach and is close to the intuitive
picture of what happens in AES.  In a way, one could view this as a
``lattice of impurities''.  Besides maintaining translational
invariance, this method has the added benefit that we can infer
apriorily the effect of removing these states, as we discuss in the
next section. Moreover, if need be one can also systematically add
some of these states back into the calculation -- for example,
starting with the states where the holes belong to neighboring TM
sites only.  Their importance can therefore be assessed
quantitatively. Generalization to systems with partially filled TM
orbitals is also less computationally costly than for the full
solution. More discussion on the meaning and relevance of this
approximation is presented below.

\section{Results}

\subsection{The periodic Anderson model}

We start by discussing the full periodic model. Besides providing the
test case against which we compare the variational and the impurity
approximations, this also allows us to understand the various features
of these spectra and their dependence on the various
parameters. Since the origin of  some of these various 
features is model independent (for example,  the location of the
continua in the AES spectrum is always obtained from the self-convolution of the
one-hole spectrum), spectra with qualitatively similar features should be expected
in more realistic models.

\begin{figure}[t]
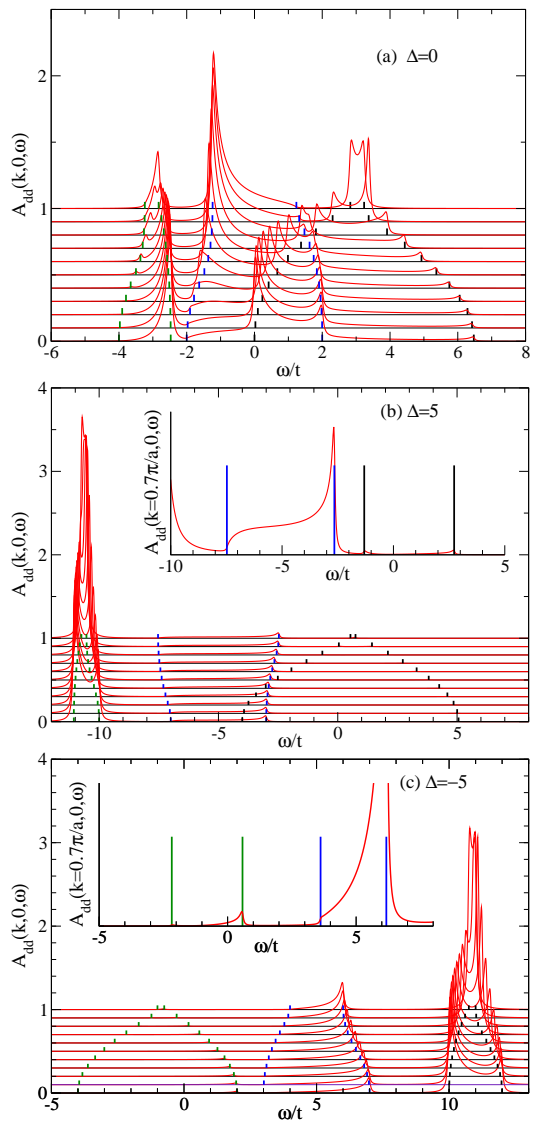

\includegraphics[width=0.8\columnwidth]{Fig2a.eps}
\includegraphics[width=0.8\columnwidth]{Fig2b.eps}
\includegraphics[width=0.8\columnwidth]{Fig2c.eps}
\caption{ (color online). AES spectral weights $A_{dd}(k,0,\omega)$
  for the full periodic 
  model vs. $\omega$ for $k\in [0,\pi/a]$. The
  curves are shifted vertically with increasing $k$. In all cases
  $t=V=1, U=U_0=0, \eta=0.05$, while (a) $\Delta=0$; (b) $\Delta =5$; (c)
  $\Delta=-5$. The small vertical lines indicated the expected
  locations of band edges. See text for more details. }
\label{fig2}
\end{figure}

To understand the dependence of the AES spectral weight
$A_{dd}(k,0,\omega)$ on the various parameters, we start by setting
$U=U_0=0$, $t=V=1$, and varying $\Delta$. Momentum-resolved results
are shown in Fig. \ref{fig2} for uniformly spaced values of $k$ from 0
to $\pi/a$.

To make sense of these fairly complicated spectra, we note that these
parameters correspond to a non-interacting system. As a result, the
two-hole spectra must be convolutions of single-hole spectra, which
are easy to calculate (note that single-hole spectra can also be
obtained experimentally from Angle-Resolved
Photoemission). Straightforward calculations show that a hole of
momentum $k$ introduced in the state $|\Omega\rangle_P$ has two
possible eigenenergies:
\begin{multline}
E_{1h}^{(\pm)}(k) ={1\over 2} \left[-U-\Delta -U_0 +2t\cos(ka)
  \right. \nonumber \\ \left. \pm
  \sqrt{\left(U+\Delta-U_0+2t\cos(ka)\right)^2 + 16 V^2 \cos^2
    {ka\over 2}}\right]. \nonumber
\end{multline}

If $\Delta$ is sufficiently large, these correspond to either having the hole
preponderantly on the TM sites (i.e., a TM-like band) or in a O-like
band. Note that the electron correlations energies $U, U_0$ enter here
because the single hole is introduced in an otherwise full band.

The convolution of these two one-hole continua result in three
two-hole continua, covering the ranges $E_{2h}^{(\gamma\delta)}(k) \in
\{E_{1h}^{(\gamma)}(k-q) + E_{1h}^{(\delta)}(q)| -{\pi\over a} < q \le
   {\pi\over a}\}$ for $\delta,\gamma =\pm$. Their band edges are
   shown by small vertical markers in Fig. \ref{fig2}; each band has a
   different color. These expected band-edges indeed agree perfectly
   with the features seen in the AES spectral weight (a broadening
   $\eta=0.05$ was used in the spectral weight, accounting for the
   apparent ``overflow'' at band edges).

For $\Delta =0$, the two upper bands overlap partially, and the
spectral weight is distributed fairly equally between all
features. For $\Delta =\pm 5$, most of the weight is in the
lowest/highest continuum which contains states with both holes in the
TM-like band and thus has the highest overlap with the state
$|k,0,dd\rangle$ of interest to AES. This band is centered at
$-2\Delta$, which is the change in energy if two holes are removed
from TM sites if there was no hybridization, $V=0$, and in the absence
of correlations $U=U_0=0$,
and is fairly narrow since the holes cannot hop directly between TM sites.
 The middle continuum, with one
hole in the TM-like band and one in the O-like band, has less weight
but is still visible. It is centered around $-\Delta$, i.e. the
energy cost for removing one electron from the TM site, and is broader
because the O-like band has significant bandwidth. The third
continuum, with both holes in the 
O-like band, has very little overlap with the state $|k,0,dd\rangle$
and is therefore only visible when the scale is significantly expanded, as shown
in the insets. It is centered around the origin (for $U_0=0$) and is
the broadest of the three features because holes can hop directly
between O sites. In the limit $V \rightarrow 0$, the maximum bandwidth
of this continuum should be $8t$, when $k=0$. We see that hopping to
and off the TM sites, controlled by $V$, has a significant
influence on this bandwidth  when $V\sim t$.   

\begin{figure}[t]
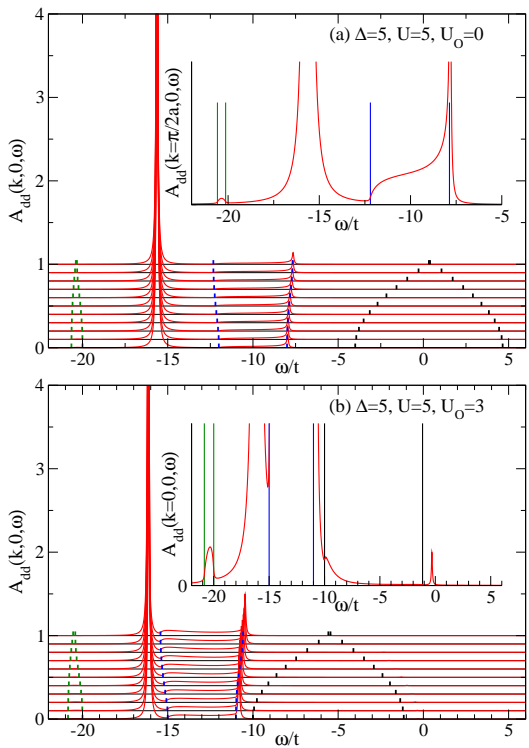

\includegraphics[width=0.8\columnwidth]{Fig3a.eps}
\includegraphics[width=0.8\columnwidth]{Fig3b.eps}
\caption{ (color online). AES spectral weights for the full periodic
  model $A_{dd}(k,0,\omega)$ vs. $\omega$ for $k\in [0,\pi/a]$. The
  curves are shifted vertically with increasing $k$. In all cases
  $t=V=1, \Delta=5, \eta=0.05$, while (a) $U=5, U_0=0$; (b) $U=5,
  U_0=3$. The small vertical lines indicated the expected 
  locations of band edges. See text for more details. }
\label{fig3}
\end{figure}

While one can set any values for parameters in a theoretical study,
physically it makes sense to focus on cases where the low-energy
states favor having the extra holes at the TM sites, so that AES is
maximally sensitive to the TM parameters. Simple arguments, listed in Appendix
B, show that this implies  $\Delta > 2t+ U_0$.

As a result, we set $\Delta =5$ and investigate the role of the
on-site correlations. Fig. \ref{fig3}(a) shows  $A_{dd}(k,0,\omega)$ for
$U=5, U_0=0$ while  Fig. \ref{fig3}(b) is for $U=5, U_0=3$. Comparing
Fig. \ref{fig3}(a) with Fig. \ref{fig2}(b), we see that the lowest
band with the two holes in the TM-like bands has shifted by about $2U$
and is now centered around $-2(\Delta +U)$, as expected since
$-(\Delta +U)$ is the energy cost for removing an electron from a
TM site (if there was no hybridization with the O bath). Similarly,
the middle band is shifted by about $U$ to around $-\Delta -U$, while
the upper band with the two holes in the O bath is essentially
unchanged.  Apart from these expected shifts, we see that most
spectral weight has moved from the lowest band (where it was for
$U=0$) into a new discrete peak that has appeared  $\sim U$ above
it. This weakly dispersing peak describes (anti)bound states with both
holes at the same TM site, hence the large overlap with
$|k,0,dd\rangle$. Indeed, it is located close to 
$-2\Delta -U$, where it would be expected to appear in the absence of
hybridization with the O bath. For our simple model, this peak
represents the ``multiplet structure'' associated with the TM
element. It will evolve into a genuine multiplet upon inclusion in
  the model of
  more orbitals at the TM site.

Correlations at the O site have similar effects, as shown in
Fig. \ref{fig3}(b). Since the cost of removing an electron from the O ion
is now lowered by $U_0$, the central band shifts by an additional $-U_0$ while the
upper band shifts by an additional $-2U_0$. On the other hand, the
lowest band with the two holes in the TM-like band is essentially
unchanged, while the ``multiplet'' peak is shifted to slightly lower
energies due to level repulsion with the central
continuum. Correlations at O sites also produce an (anti)-bound
discrete state with both holes at the same O site which is pushed above
the upper continuum. However, it has
very little weight and can only be seen on a greatly expanded scale,
as shown in the inset, as a small peak located just above the
highest-energy band edge.

To summarize, the location of the continua in the momentum-resolved
AES spectrum must agree with the self-convolution of the one-hole spectrum,
which can be obtained from ARPES. Additional discrete peaks (or strong
sharp resonances, if these happen to fall inside another continuum)
indicate the presence of on-site correlations, and allow one to find
various 
on-site parameters from the corresponding multiplet
structure.\cite{Antonides}
 Of course, because the projection is on the $|k,0,dd\rangle$ state, the
spectral weight is large for states which predominantly have both
holes at the same TM site. As discussed in the previous section, it is
trivial with our method to project on other states, for example with
both holes at the same O site. While the spectrum is unchanged, this
will shift the spectral weight between the different
features.

\subsection{Variational approximation}

\begin{figure}[t]
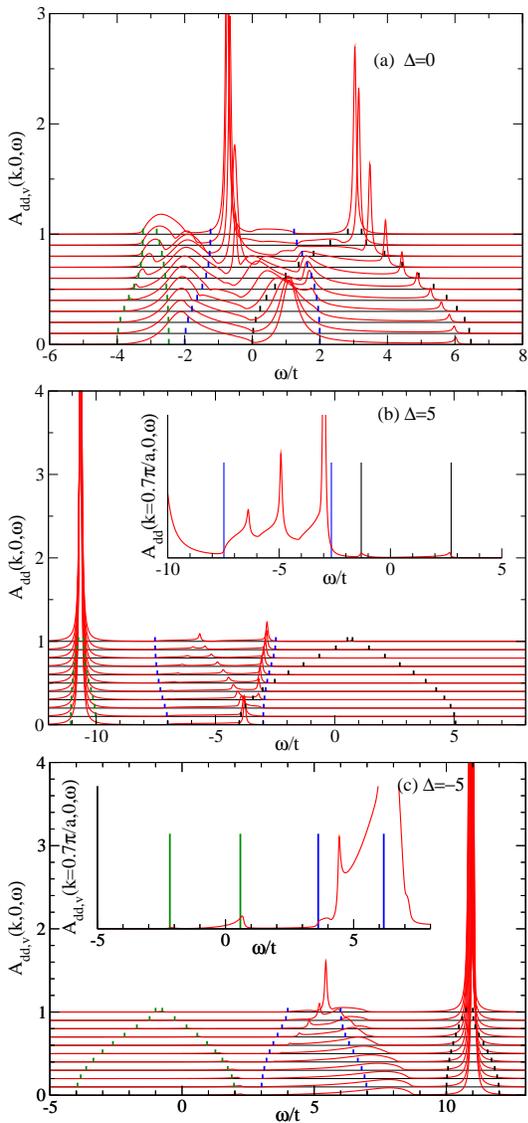

\includegraphics[width=0.8\columnwidth]{Fig4a.eps}
\includegraphics[width=0.8\columnwidth]{Fig4b.eps}
\includegraphics[width=0.8\columnwidth]{Fig4c.eps}
\caption{ (color online). Same as in Fig. \ref{fig2}, but  the AES
  spectral weight is obtained with the
  variational approximation. }
\label{fig4}
\end{figure}

\begin{figure}[t]
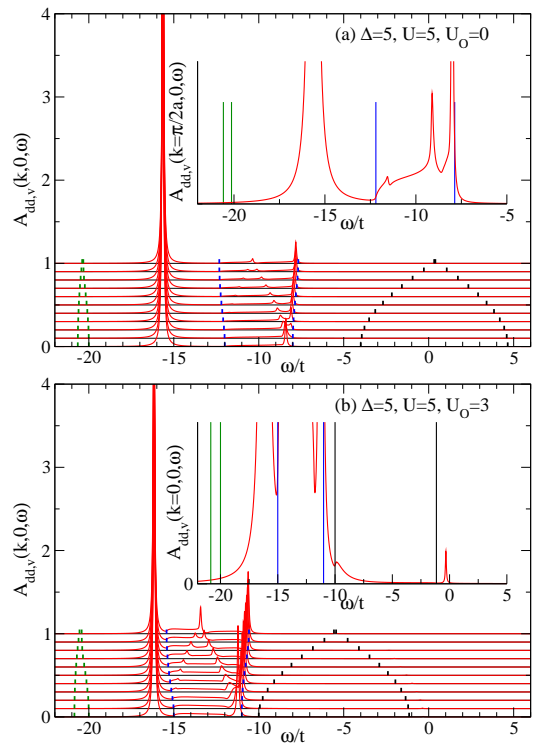

\includegraphics[width=0.8\columnwidth]{Fig5a.eps}
\includegraphics[width=0.8\columnwidth]{Fig5b.eps}
\caption{ (color online). Same as in Fig. \ref{fig3}, but  the AES
  spectral weight is obtained with the
  variational approximation.   }
\label{fig5}
\end{figure}

In Figs. \ref{fig4}
and \ref{fig5} we present data obtained with our variational
approximation for the same sets of parameters. As before, the  
vertical markers show the expected location of the band-edges, based
on the convolution of the one-hole spectra.

Starting with the un-correlated case in Fig. \ref{fig4}, we see that
this variational approximation is very poor when $\Delta=0$, in panel
(a). While the overall spectral range is in fair agreement with that
predicted by the two-hole convolutions, inside this interval there is
significant disagreement. In particular, the variational calculation
predicts spectral weight at energies that should be gaped. This
disagreement is not surprising, since in this case we are projecting
out states that have energies very similar to the states we keep, and
this is not a sensible strategy.  

As shown in panels (b) and (c), the situation improves when $\Delta$
is large enough to separate a TM-like band from the O-like band. In this
case, the agreement for the two less visible bands is reasonable, even
though the projection may slightly over/underestimate the location of
the band-edges and it produces additional structure within
the bands, especially the central one. This is not surprising since one
expects a fair amount of hybridization between states in the central
continuum, which have one hole in the TM-like band and one in the
O-like band, with the states with the two holes at different TM sites
that are projected out.  

Finally, the continuum with the two holes in the TM-like bands is
replaced by discrete peaks located at roughly the correct energy. The
disappearance of this continuum is  expected, since we
projected out precisely the states responsible for generating it,
i.e. with holes at different TM sites. In the absence of correlations,
the state with both holes at the same TM site has a similar energy,
hence the peak. Similar conclusions hold for $\Delta = -5$. 

If we turn on the correlations, this ``multiplet structure'' peak
moves away from the continuum. Indeed, as shown in both panels of Fig. \ref{fig5},
now the low-energy continuum with the two holes in the TM-like band is
completely absent from the variational results but the important
feature, i.e. the high-weight ``multiplet structure'' peak, is
essentially indistinguishable from that predicted by the full
solution. The higher energy continua show some additional structure
within this variational calculation, but since this structure also appears in the
absence of correlations there is little danger of confusing it
with real features induced by the correlations.

To summarize, for parameters likely relevant for the real materials,
i.e. when the TM-like and O-like bands are sufficiently well
separated, the variational approximation does a very good job in
capturing the multiplet structure due to TM on-site correlations. One
of the continua is projected out, but it has little weight and its
location is apriorily known from the one-hole spectrum. The higher
energy continua are reproduced at roughly the correct locations but
with some additional features, which however also appear in the absence of
correlations. With some care, one can use this more efficient
calculation to understand most, if not all, the important physics
contained within these spectra.

\subsection{Impurity approximation}

For comparison, we also present AES spectra obtained with the impurity
approximation. Since here the translational invariance is lost, we can
only show one $A_{dd}(\omega)$ curve for each set of parameters. This
can be roughly thought of as a momentum-integrated spectral weight. 

\begin{figure}[t]
\includegraphics[width=0.6\columnwidth]{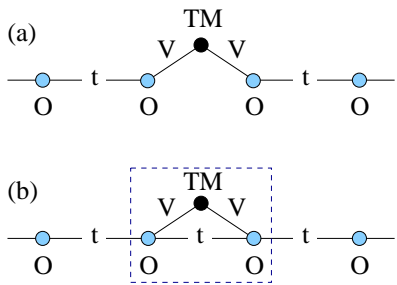}
\caption{  Two  ways to connect the
  impurity to the O bath: (a) trivial, and (b) non-trivial topology. The latter
  can be reduced to the former if one considers the three central
ions enclosed by the box to form an effective ``impurity''.}
\label{fig6}
\end{figure}

\textit{1. Trivial topology:} We begin with the simpler
impurity model sketched in Fig. \ref{fig6}(a), where the TM impurity is connected to two
independent semi-infinite half chains.  

If $\Delta$ is sufficiently large, the single-hole spectrum  consists
of a discrete peak (hole localized at TM site) 
and a continuum (hole in the O-band). 
As a result, the 2-holes convolution  consists of three features: (i)
a discrete state with both holes at the TM 
site (the ``multiplet''); (ii) a continuum of states with one hole occupying the
TM impurity state and the other moving freely in the O bath; and (iii) the
two-hole continuum with both holes in the O bath, which here spans
the interval $[-2U_0-4t, -2U_0+4t]$ irrespective of the value of
$V$. 

\begin{figure}[b]
\includegraphics[width=\columnwidth]{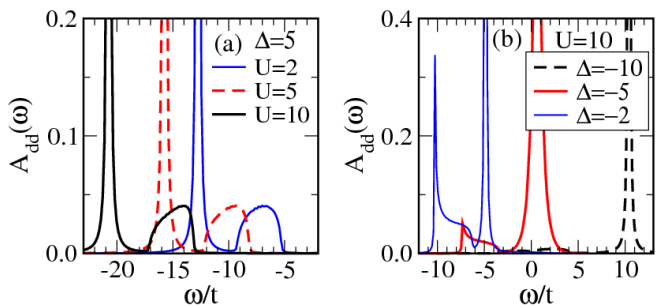}
\caption{ (color online). Impurity spectral function $A_{dd}(\omega)$
  vs. $\omega$ for the trivial topology of Fig. \ref{fig6}(a) with
  $t=V=1$ and  (a) $\Delta=5, U=2,5,10, U_0=0$, and (b) $\Delta=-10,
  -5, 2, U=5, U_0=0$.} 
\vspace{-.0cm}
\label{fig7}
\end{figure}

Results shown in Fig. \ref{fig7} for several sets of parameters
 indeed show these features, although the two-hole continuum
can only be observed on a magnified scale.

\begin{figure}[t]
\includegraphics[width=\columnwidth]{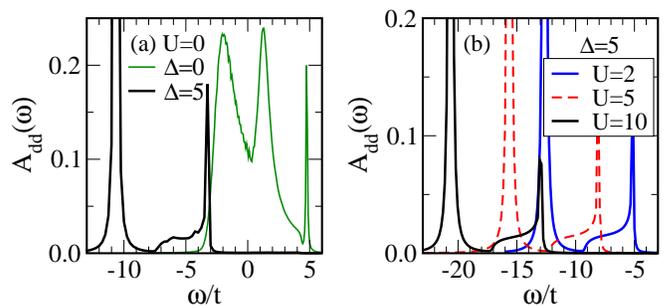}
\caption{ (color online). Impurity spectral function
  $A_{dd}(\omega)$ vs. $\omega$ for the non-trivial
  topology of Fig. \ref{fig6}(b) with $t=V=1$ and (a) $\Delta=0, 5$
  and $U=U_0=0$,  (b)  $\Delta=5, U=2,5,10, U_0=0$.  }
\vspace{-.0cm}
\label{fig8}
\end{figure}

\textit{2. Non trivial topology:} Results for similar parameters, but
for the non-trivial topology of Fig. \ref{fig6}(b), are shown in
Fig. \ref{fig8}. Here the impurity forms a ring with its two neighbour
O sites. This non-trivial topology can support more impurity states
localized on the ring
than the trivial topology discussed above,  so one may expect a more
complex two-hole spectrum.

In panel (a), we plot $A_{dd}(\omega)$ for $t=V=1, U=U_0=0$ and
$\Delta=0,5$. The $\Delta=0$ spectrum shows two broad resonances in
the two-hole continuum plus a discrete peak above it. These are
understood in terms of the ring ``multiplet''. For these parameters,
the isolated ring has three two-hole eigenstates at $-2t, t, 4t$. The
two broad resonances are associated with the former two, while the
latter falls at the upper edge of the continuum and is pushed above it
by level repulsion. States with one hole on the ring and the other in
the bath overlap with the two-hole continuum.  For $\Delta=5$ (thick
black line) there are two one-hole ring eigenstates at $-5.32t,
1.32t$, resulting in three two-hole ring eigenstates at $-10.63t, -4t,
2.63t$. The latter two fall inside the two-hole continuum, so only the
former is visible as a discrete peak. The continua with one hole on
the ring and one in the bath span $[-7.32t, -3.32t]$ and $[-0.68t,
  3.32t]$, respectively. Only the former is (partially) distinct from
the two-hole continuum, and is visible in the spectrum. Very little
weight is left in the two-hole continuum, which is not visible on this
scale. The effect correlations, shown in panel (b),
can be explained with similar arguments.

Interestingly, these results are qualitatively like those of
Fig.~\ref{fig7} (a) for the 
trivial topology, because the additional
states in the ring multiplet fall inside the two-hole continuum and
are washed out into broad, low-weight resonances. This is certainly
true for $U\gg t, |\Delta|\gg t$, when the TM states become (to zero
order in perturbation theory) eigenstates of the ring, with the
remaining ring eigenstates involving only O sites and therefore being
located inside the two-hole continuum.   

 \begin{figure}[b]
\includegraphics[width=\columnwidth]{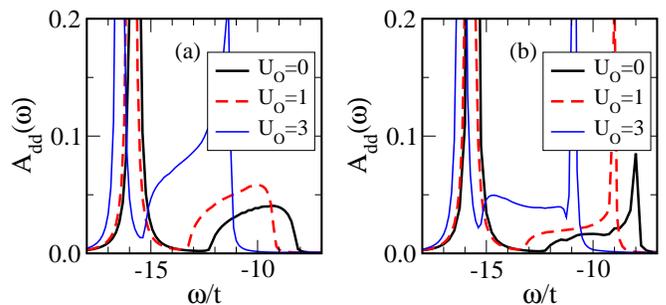}
 \vspace{-.0cm}
 \caption{(color online). Impurity spectral function
  $A_{d\uparrow,d\downarrow}(\omega)$ vs. $\omega$ for (a) the trivial
  topology and (b) the ring
  topology,  with $t=V=1, \Delta=U=5$ and $U_0=0,1,3$.}
 \vspace{-.0cm}
 \label{fig9}
 \end{figure}

\textit{3. Bath correlation effects:} In Fig. \ref{fig9} we briefly
show the effect of adding on-site repulsion at the O sites in the
impurity approximation. Panel (a) is for the trivial topology while
panel (b) is for the ring topology.  The main effect of $U_0$ is to
shift the continuum with one 
hole in the bath
to lower energies by $U_0$, as expected. This leads to a stronger
hybridization with the discrete peak, which is consequently pushed to
lower energies as well. The weights in these features also vary, but
the two-hole continuum remains invisible on 
this scale.  

\begin{figure}[t]
  \centering{ \includegraphics[width=0.9\columnwidth]{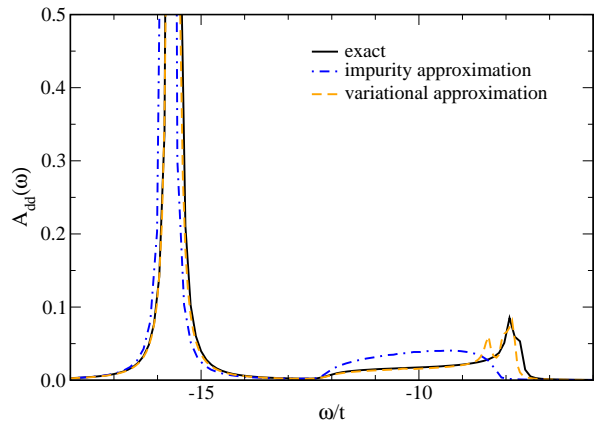}}
  \caption{Momentum integrated AES spectral weight ${1\over
      N}\sum_{k}^{} A_{dd}(k,\omega)$ for the exact solution (full
    line) and our variational approximation (dashed line) compared to the
    prediction of the impurity approximation with trivial topology
    (dot-dashed line), for $t=V=1, U=\Delta=5, U_0=0$.}
\label{fig10n}
\end{figure}

To summarize, we find that the results for the two topologies are
rather similar if $|\Delta|$ and/or $U$ are large compared to $t,V$
(if this is not the case,  the impurity approximation is 
not valid). 
In this limit,  the  impurity approximation is quite successful in
reproducing the location of the multiplet peak, as shown in
Fig. \ref{fig10n} where we compare $A_{dd}(\omega)$ for the trivial
topology (dot-dashed line) against ${1\over N}\sum_{k}^{} 
A_{dd}(k,\omega)$  
  for the exact solution (full line) and our variational approximation
  (dashed line), for  $\Delta = U=5, U_0=0$. The multiplet
    peak   disperses very weakly for these parameters so all three
    curves show 
  what looks like a discrete peak at low energies. As expected, the variational
  approximation provides a  more accurate estimate of the peak
  position.  The impurity approximation is also less successful in
  predicting the  correct bandwidth for the continuum visible at
  higher energies; when analyzing   experimental data, this  may
  result in wrong values assigned  to the hopping
  parameters. Fig. \ref{fig10n} confirms  that our variational 
  approximation is more accurate than the impurity  approximation.  

\section{Discussion $\&$ Conclusions}

In this work we introduced a novel variational approach to calculate
two-hole Green's functions needed for AES spectroscopy, and 
applied it to a simple 1D model with otherwise full bands, so that we
can compare its predictions against the exact solution available for
this case. We also calculated the results in the impurity
approximation, for two different ways of connecting the impurity to
the O bath.

Both approximations are reasonable to use only if the single-electron
parameters are such that one-hole eigenstates with the hole primarily
at the TM site (i.e., of TM character) are energetically well
separated from those with the hole primarily in the O band. If this is
not true, then the basis states projected out in these approximations
are energetically close to those kept within the calculation, and the
results are not sensible.

If the above mentioned condition is satisfied, our variational
approximation is superior to the impurity 
approximation. Not only does it produce momentum-resolved results, but
the location of the continua that appear in the AES spectrum is in
fair agreement with that expected from the convolution of the one-hole
spectra, unlike for the impurity approximation.  The superiority
  of our method can also be justified as follows: both methods are
  variational, since both limit the basis of allowed two-hole states.
  Our method has a much bigger variational space, therefore it has to
  be more accurate. The results presented here confirm it.

Since the one-hole spectrum can be obtained from ARPES measurements
(and, from a theoretical perspective, can be calculated with methods
analogous to those we use here to find the two-hole propagator), a
combination of the two spectroscopies can be used to determine which
features in the two-hole spectrum come from this convolution. Any
other features must be due to on-site correlations, and can therefore
help pinpoint the values of various on-site interaction energies.

The method of Ref. \onlinecite{cf-prl} can be extended
straightforwardly to compute Green's functions for problems with three
or more holes. As a result, it can certainly be used to solve
  impurity-type problems where the TM has a partially filled $d$ shell
  while the O band is completely full prior to the Auger process. If
  the TM has a $3d^n, n < 10,$ configuration, then after the Auger
  process there are $12-n$ holes in the impurity problem. While
  computational times depend on $n$ and on how complicated the O bath
  is (how many $2p$ orbitals per O, and in what dimension), at least
  some of the higher $n$ cases should be solvable exactly. One can
  make further progress using the fact that processes where very many
  more than 2 holes hop into the O bath should be energetically very
  costly (else, the ground-state would have partially filled O bands
  to begin with). As a result, one could systematically increase the
  number of holes allowed to hop into the bath starting from 2, until
  convergence is reached. This may allow one to investigate all
  possible $n$ values within the impurity approximation, if
  convergence is reached fast enough.

Similar considerations apply to our variational method, because the
total number of basis states it includes (for each fixed value of $k$)
is smaller but comparable to that for the impurity approximation. As a
result, one should get more accurate and momentum-resolved results at
comparable computational costs.

Of course, not all oxides have full O $2p$ bands, although many do for
at least some particular doping (\textit{e.g.}, an insulating parent
compound). We envision using this method to extract information about
crystal field splitting and on-site and nearest-neighbor interactions
from Auger spectroscopy on 
this compound. If further doping that results in partially
filled O $2p$ bands does not result in significant additional
screening of these short-range interactions, then these values would
be relevant over the entire doping range.

In conclusion, we believe that this method proposes an efficient way to make progress on
understanding AES spectra for systems where the $d$ orbitals of the TM
are only partially filled, so long as the states of TM character are
not too close to the filled O states. Such work is now in progress.

\begin{acknowledgments} This work was supported by NSERC, CIfAR and QMI. 
\end{acknowledgments}

\appendix

\section{Continued-fraction solution for the impurity approximation}

\begin{figure}[t]
  \centering{ \includegraphics[width=0.7\columnwidth]{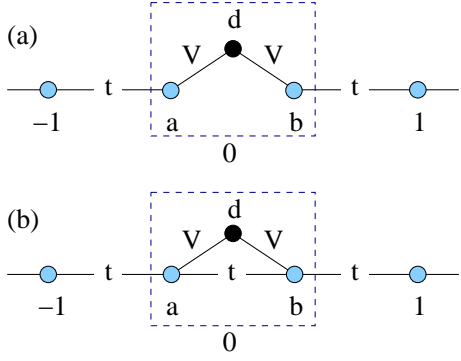}}
  \caption{ Actual indexing of the O sites used in this work. The
    three central atoms comprise 
    ``site'' 0.}
\label{fig22}
\end{figure}

For technical reasons which will become apparent soon, it is more
convenient to group the TM together with the two O sites it directly
hybridizes with into ``site'' 0, and to index the O atoms to its
right/left as $\pm 1, \pm 2$, etc. The two O inside ``site'' 0 will be
called ``a'' and ``b'', respectively. This indexing is shown in
Fig. \ref{fig22}. Amongst other things, this indexing makes it easy to
study the two situations depicted in Figs. \ref{fig22}(a) and (b),
where the hopping between the two central O is turned off/on.

We use the operator identity $(\omega+ E^I_\Omega+i\eta-{\cal
  H_I})\hat{G}_I(\omega)=\hat{I}$ to generate EOMs for the
Green's functions. For example, for any $|i| >1, |j| >1$, this results
in: $(\omega+E^I_\Omega+U_0 (2-\delta_{i,j})+i\eta)G_{i{\uparrow},
  {j\downarrow}}(\omega)= tG_{{i\uparrow}, {j+1\downarrow}}(\omega)+
tG_{{i-1\uparrow}, {j\downarrow}}(\omega)+ tG_{{i\uparrow},
  {j-1\downarrow}}(\omega)+ tG_{{i+1\uparrow},
  {j\downarrow}}(\omega)$, where $G_{i\sigma,
  {j\sigma'}}(\omega)=\langle
dd|\hat{G}_{I}(\omega)|i\sigma,j\sigma'\rangle$. In other words, it
links together 
propagators where the holes are at a distance $|i-j|=n$ apart, to
propagators where the holes are $n\pm 1$ sites apart. This  allows us
to rewrite these equations of motion as 
simple recurrence relations by grouping together in a vector $V_n$ all
the propagators where the holes are $n$ sites apart. Note that this is
true for states with a hole at ``site'' 0 and one at any site $|n|\ge
1$, which also enter into $V_n$ and only link to propagators in
$V_{n\pm 1}$.
 
To be more precise, let:
$$V^{\sigma, \sigma'}_n= \left( \begin{array}{c}
  .\\.\\.\\G_{{i\sigma}; {i+n,\sigma'}}(\omega) \\G_{{i-1,\sigma};
    {i+n-1,\sigma'}}(\omega) \\ G_{{i-2,\sigma};
    {i+n-2,\sigma'}}(\omega)\\.\\.\\.
\end{array} \right)$$
be a vector of infinite length which contains all the propagators with
the holes $n$ sites apart. Of course, there are three entries
replacing the $(0,n)$ entry, corresponding to sites $a,b,d$, and
similarly for $(-n,0)$.  Further, we note that for $n\ge 1$, the
ordering of the spins is preserved under hopping, since the left-most
hole cannot pass by the right-most hole with one hop.  Thus, we can
write a recurrence relation for these vectors, for any $n\ge 1$, as:
\begin{multline}
\left(\begin{array}{cc} \gamma_{n} & 0\\ 0 &
  \gamma_{n}\end{array}\right)\left(\begin{array}{c}
  V_{n}^{\uparrow\downarrow}\\ V_{n}^{\downarrow\uparrow}\end{array}\right)=\left(\begin{array}{cc}
  \alpha_{n} & 0\\ 0 &
  \alpha_{n}\end{array}\right)\left(\begin{array}{c}
  V_{n-1}^{\uparrow\downarrow}\\ V_{n-1}^{\downarrow\uparrow}\end{array}\right)\\ +\left(\begin{array}{cc}
  \beta_{n} & 0\\ 0 &
  \beta_{n}\end{array}\right)\left(\begin{array}{c}
  V_{n+1}^{\uparrow\downarrow}\\ V_{n+1}^{\downarrow\uparrow}\end{array}\right)
\end{multline}
where $\gamma_n$, $\alpha_n$ and $\beta_n$ are very sparse matrices,
whose elements can be read off from the equations of motion as in the
periodic case. Further, as before, the tridiagonal form of the
recurrence relation admits a continued fraction solution for these
vectors. The rest of the procedure is the same as for the periodic
case.

\textit{Truncation schemes of continued fractions:}

(i) For this method to work we need to truncate
the continued fraction at a large inter-hole separation $M$, both
for the periodic and impurity cases.  In the periodic system, we start
out by creating two holes on a TM site.  The two holes can now
delocalize in the system. However, the broadening $\eta$ introduces a
finite lifetime ($\sim 1/\eta$) so that
$<k,0,dd|{G}_P(\omega)|k,n,\alpha\beta>\longrightarrow0$
as $n\longrightarrow\infty$. Thus for a large
enough $n=M$, we can truncate the continued fraction for the
V-vectors by setting $V_{M_c+1}$ to zero.

For the impurity problem, this is justified because $G_{i\sigma; i+n,
  \sigma'}(\omega)$ is the Fourier transform of the amplitude of
probability that the two holes move from the impurity (TM) site to the
sites $i, i+n$ within a time $\tau$. The larger $n$ is, the less
likely this process becomes, hence the smaller these propagators must
be. This is certainly true for the energy ranges spanned by
eigenstates that favor having the holes on the impurity (TM) site,
since then they are unlikely to wander very far away from it. However,
this is also true even if the holes preferred the O bath. This is
because the broadening $\eta$ is equivalent to introducing a finite
lifetime $\sim 1/\eta$ for the holes, so they  cannot move
arbitrarily far in a finite time $\tau$. Of course, in
this latter case the appropriate cutoff $M$ increases as $\eta$
decreases. In practice, for both cases, we increase $M$ until the
results become insensitive to further changes.

(ii) A further truncation is necessary for the impurity problem. This
is because, with the ordering of the propagators used for the $V_n$
vectors, we have the additional complication that all these vectors,
and therefore all the sparse matrices $\alpha_n, \beta_n, \gamma_n$,
are infinitely-dimensional. In order to calculate the continued
fractions, we need to truncate the size of these vectors, as well. The
reasons discussed above justify doing this if either $|i|\gg 1$ and/or
$ |i+n|\gg 1$. We use the following truncation procedure: for a fixed
separation $n$ between the holes, the maximum distance either of the
two holes is allowed to travel away from the TM is $R_c+n$,
\textit{i.e.}, we truncate $V_n^{\sigma,\sigma'}$ at $G_{{R_c,\sigma};
  {R_c+n,\sigma'}}(\omega)$ at the top and $G_{{-R_c-n+1;\sigma};
  {-R_c+1,\sigma'}}(\omega)$ at the bottom. Again, $R_c$ is increased
until results become independent of its value. As a final comment, we
note that there are other ways of grouping the propagators into
vectors so that the equations of motion still lead to a simple
recursive relation. Different schemes have various computational
advantages and disadvantages, but they all convergence to the correct
answer if the cutoffs are sufficiently large. This converged result is 
equivalent to the exact solution computed by the Cini-Sawatzky theory for the Anderson impurity problem.

\section{Choice of parameters} 

 For
simplicity, we discuss the impurity approximation first, and the full
periodic system second.

For the impurity approximation, if we ignore the TM-bath
hybridization, $V\rightarrow 0$, then if both holes are at the TM site
the energy of the state is $E^I_\Omega - 2\Delta -U$; if one hole is
at the TM site and one is in the O bath, the minimum energy of such
states is $E^I_\Omega - \Delta -U -2t-U_0$. Finally, if both holes are
in the O bath, the minimum energy of such states is $E^I_\Omega - 4t -
2 U_0$. As a result, the state with both holes at the TM atom is
favorable energetically if $\Delta > 2t +U_0$ and $2\Delta + U > 4t +
2 U_0$. The second condition is automatically satisfied if the first
one holds, since the TM on-site interaction is repulsive, $U>0$.

The condition $\Delta > 2t +U_0$ implies that AES should  be more useful
for Mott insulators than for charge transfer materials.\cite{ZSA}
To see this, consider one-electron removal states from the GS. If the
hole is removed from the TM site, the energy of the state is
$E^I_\Omega - \Delta -U$, while if it is removed from an O site, the
minimum energy is $E^I_\Omega - 2t - U_0$. The latter is energetically
more expensive than the former if $\Delta > 2t +U_0$. This makes
sense, because if the material was a charge transfer insulator the
additional holes would prefer to stay in the O bath and AES would be
less sensitive to the TM atom specific properties.

The analysis for the periodic system is very similar for
$V\rightarrow 0$, because the energy differences between states having
(i) both holes at the same TM site, (ii) one hole at a TM site and one
in the O bath, and (iii) both holes in the O bath, are precisely the
same as for the impurity case. Because of this and also in order to be
able to meaningfully compare with impurity approximation results,
we use the same parameters in both cases. 

Note that in the periodic system we 
can also have (iv) the two holes at different TM sites. For $V=0$, these states has
energy $E^P_\Omega - 2\Delta -2U$ and are always energetically favored
compared to having both holes at the same TM site. Of course, these
are the states projected out in our variational calculation.

\end{document}